# Gavial: Programming the web with multi-tier FRP

Bob Reynders[a], Frank Piessens[b], and Dominique Devriese[c]

a   Chonnam National University
b   imec - DistriNet - KU Leuven
c   Vrije Universiteit Brussel

**Abstract**   Web applications are inherently distributed, and not just because their client and server counter-parts run on networked systems. Web applications are written in multiple programming languages and as multiple programs: the server and client programs. In an effort to lower the complexity of the web, multi-tier programming was proposed. In multi-tier programming languages the language and its tooling give support to create web applications as a whole, one program is written in one language.

Web applications are also inherently asynchronous. On the server, they constantly process several client requests and in the browser they constantly have to react to input, be it from the user or a server. A technique that can be applied to make such programs easier to understand is functional reactive programming. A functional programming model that models an interactive program as compositions between two primitives, behaviors and events.

Developing web applications requires dealing with their distributed nature and the natural asynchronicity of user input and network communication. For facilitating this, different researchers have explored the combination of a multi-tier programming language and functional reactive programming. However, existing proposals take his approach only part of the way. Some parts of the application remain imperative, do not consider network traffic overhead of incremental data, compatibility with common APIs like XMLHttpRequest etc., or they do not consider glitches across network communication (or do but introduce locking or overhead)

In this paper, we present Gavial: a design and practical implementation of multi-tier FRP that allows constructing a web application as a functionally reactive program. We propose a novel integration of existing techniques such as solutions to bootstrapping web applications, recursive behaviors for web interfaces, asynchronous FRP and incremental behaviors. As well as novel support for *tiered glitch freedom*, a compromise between performance and glitch-free evaluation of the distributed FRP program, and a three-tiered system suitable for request-response-style applications as well as user-to-user applications. Gavial automatically runs in request-response (XMLHttpRequest) mode or in websocket mode.

After applying a number of old and new ideas to create a practical multi-tier FRP implementation in Gavial, we demonstrate it by building an example and show that it can in fact deal realistically with important practical aspects of building web applications. At the same time, we retain the declarative nature of FRP, where behaviors and events have an intuitive, compositional semantics and a clear dependency structure.

## The Art, Science, and Engineering of Programming

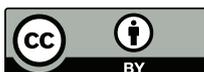

## 1   Introduction

Developing web applications requires dealing with the specificities of the web. This includes the distributed nature of applications, partly executing on the client (i.e. the user's browser), partly on the server, and the different parts communicating over APIs like WebSockets or XMLHttpRequests. Additionally, user input and client-server communication on the web are both naturally asynchronous. These characteristics of web applications have led researchers to design programming languages or frameworks tailored to their development.

Two ideas to cope with the web's distributed and asynchronous nature are *multi-tier programming languages* and *functional reactive programming*. Multi-tier languages (see among others: [8, 19, 29]) allow both client and server parts of a web application to be written in a single codebase — offering a joint semantics and allowing cross-tier abstractions. On the other hand, functional reactive programming [12] is an alternative programming model that facilitates development of asynchronous applications and reasoning about their behavior. Instead of using side-effecting callbacks, FRP programs are constructed by composing *behaviors* and *events*: components representing time-dependent values.

Combinations of FRP with multi-tier programming have been explored in the past. Multi-tier FRP has been presented with Scala in [25]. They provide .to(`Client`/`Server`) on events and behaviors. Eliom [23] provides a client/server reactive abstraction (since v5.0).[1] They also provide a *signal* and have similar tier-crossing primitives. These approaches conveniently connect multiple FRP programs across tiers, however they do not provide a unified programming experience. Programmers have to pay close attention not to get partial propagation (glitches) throughout the distributed FRP program in order to maintain consistency — something reactive programming typically takes care of.

Distributed reactive programming languages [11, 16, 18] have distributed consistency guarantees and allow distributed FRP propagation without observable partial propagations (glitches). ScalaLoci [35] in particular — a general purpose multi-tier extension of Scala which allows programmers to define their own distribution scheme — supports tier-crossing primitives with distributed consistency guarantees. These guarantees are gained through a compromise of using global coordination or increasing message count or size.

The main contribution of this paper is Gavial: a design and implementation of a multi-tier FRP framework that allows constructing a web application as a functional reactive program: GUI definitions, request/response-style web apps as well as collaborative web apps such as video games are compositions of FRP primitives. We achieve this by incorporating (1) existing ideas: automatic bootstrapping [25], asynchronous FRP [9], recursive behaviors for building web interfaces [26] and incremental behaviors [24] and (2) introducing novel ideas: a form of glitch minimization across network communication called tiered glitch freedom (section 3.6), the three-tier structure of

---

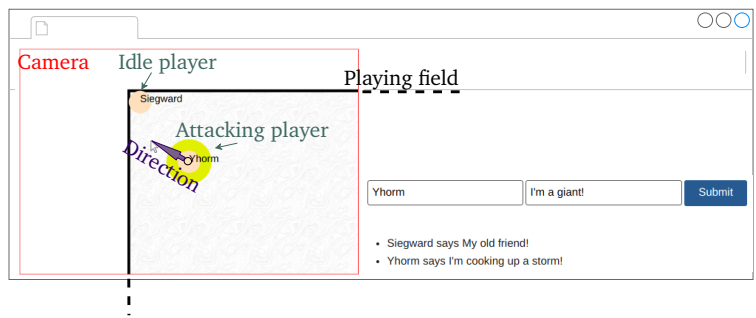

**■ Figure 1** CircleRoyale

our behaviors and events (section 3.2), novel support for using XMLHttpRequests (instead of WebSockets) in the absence of server-initiated tier-crossing (section 3.7), and an implementation reusing existing Scala infrastructure and library eco-system (section 4). Gavial shows that multi-tier FRP for the web requires only thin abstractions on top of proven technologies. At the same time, we retain the declarative nature of FRP, where behaviors and events have an intuitive, compositional semantics and a clear dependency structure.

**Outline**    First, we gradually introduce Gavial and its features by incrementally developing a game CircleRoyale (section 2). Next, we provide more details on Gavial APIs (section 3) and our implementation (section 4). We discuss related work in section 5 and conclude in section 6.

## 2    Multi-tier FRP by Example

Gavial is an embedded domain specific language in Scala (JVM) and Scala.JS (a mature compiler targeting JavaScript). We introduce it using "CircleRoyale": a small game shown in figure 1. It is inspired by a trend of simple online multiplayer games such as `agar.io` where players battle each other on a large playing field with minimal controls. In CircleRoyale, players (shown as circles) continuously move around in the direction of the mouse cursor. They can start attacks by hitting space, spawning a larger flashing circle around the player. It remains active for 2 seconds and then cannot be used for 3 seconds (the cooldown period). The game ends when the player is hit by an attack and the end score is the time spent alive.

In this chapter, we demonstrate this by gradually implementing CircleRoyale, introducing and highlighting features of Gavial along the way. We start from a simple single-player single-tier UI, and make small, local and understandable changes towards the end result. All code examples are valid Scala and actually running code. Example code is shown inline and detailed API information in captioned listings. We encourage the reader to try out Gavial on http://tzbob.be/gavial, only one command is required to setup a basic project and start the tutorial . Both the inline example (in stages) and the CircleRoyale game are available online: https://github.com/tzbob/circleroyale.

■ **Listing 1** Event and DBehavior

```
trait Event[A] { def map(f: A ⇒ B): Event
                 def fold(init: B)(f: (B, A) ⇒ B): DBehavior
                 def hold(init: A) : DBehavior[A] }
trait DBehavior[A] {
  def changes: Event[A]
  def map(f: A ⇒ B): DBehavior
  def map2[B, C](db: DBehavior)(f: (A, B) ⇒ C): DBehavior[C]
  def snapshotWith[B, C](ev: Event)(f: (A, B) ⇒ C): Event[C]
  def sampledBy(ev: Event[_]): Event[A]
  def snapshotWith[B, C](
    other: DBehavior)(f: (A, B) ⇒ C): DBehavior[C]
}
object DBehavior { def constant[A](a: A): DBehavior[A] }
```

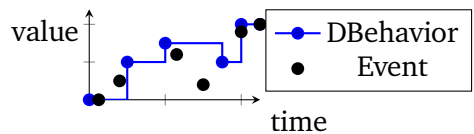

■ **Figure 2** FRP Primitives

### 2.1 Client Prototype

We start with a working solo-player version of CircleRoyale in which a player can move around and start attacks but there are no opponents.

**Functional Reactive Programming** User controls are modeled as *direction* and *attacking*:

```
val attacking: DBehavior^C[Boolean] = -- cut --
val direction: DBehavior^C[Vec2D] = -- cut --
```

To understand this, let us clarify some terminology. Gavial offers a flavor of FRP that contains: events, behaviors, discrete behaviors and incremental behaviors which we discuss later on. As depicted in figure 2, events are streams of timestamped values and behaviors are time-varying values. Discrete behaviors only change at discrete times, which behave as right continuous step functions. For now, we use only `Event` and `DBehavior`, the superscript $^C$ refers to computations on the *client* (see below for other tiers). Events and behaviors have the API shown in listing 1.

Events can be mapped over with a function or be folded into discrete behaviors. Such folded discrete behaviors start out with the given initial value and "step" to a new value when the event fires, combining the event's value and the behavior's previous value using f. `hold` is like `fold` but simply stores the last event value seen.

Behaviors can be combined using `map2` and can be read at the rate of an event using `snapshotWith` (using a function to combine the values). `sampledBy` is like `snapshotWith` but ignores the value of event `ev`. A `snapshotWith` for discrete behaviors works like `map2` but the resulting behavior changes only when other does.

■ **Listing 2** Delayed DBehavior

```
object DBehavior { def delayed[A](db: ⇒ DBehavior[A]): Behavior[A] }
```

Returning to CircleRoyale, a type `Player` represents a player's position, and whether he is alive, attacking or dead. The `svg` method produces the player's circle and optionally the attacking circle as SVG elements.

```
case class Player(position: Vec2D, attacking: Boolean, dead: Boolean) {
  def update(direction: Vec2D, attacking: Boolean): Player = -- cut --
  def setDead(dead: Boolean): Player = -- cut --
  val svg: UI.HTML = -- cut -- }
```

We define the player's state over time, by tupling the user controls (direction and whether an attack should be started) using `map2`[2] and then `folding` these changes starting from a default state. For this example we simplify a bit and allow players to attack whenever and for however long they want.

```
val input: DBehavior^c[(Vec2D, Boolean)] = direction.map2(attacking){ (_, _) }
val player: DBehavior^c[Player] =
  input.changes.fold(Player.default) { case (p, (dir, att)) ⇒ p.update(dir, att) }
```

**Throttling the update rate**  The user input in `direction` is based on mouse movement and updates very frequently, so for efficiency, we change `player` to throttle the rate. This is easy to do in Gavial using `IntervalCycle` (an abstraction around JavaScript's `setInterval`). We sample `input` by an event `time` which fires at 10 Hz and update `player` to use this throttled input instead.

```
val time: Event^c[Time] = new IntervalCycle(1.second / 10).elapsedTime
val throttledInput: Event^c[(Vec2D, Boolean)] = input.sampledBy(time)
val player: DBehavior^c[Player] =
  throttledInput.fold(Player.default) { case (p, (dir, att)) ⇒ p.update(dir, att) }
```

**Recursive Behavior**  As pointed out before [26], user interfaces are often inherently recursive. For example, in CircleRoyale the future direction of a player is relative to its current position, which is itself determined by the direction of movement. Our flavor of FRP permits explicit recursive definitions (in Scala) through `.delayed` as shown in listing 2.

A delayed behavior will have the same step function as the original behavior but it is *left* continuous instead of being right continuous as in figure 2. In other words, the delayed version of a discrete behavior keeps the old value for an instant longer when a change occurs. Note that the `db` parameter is passed by-name and is not evaluated immediately, this Scala feature (and because of the implementation of `delayed` that makes use of it) makes it possible to make *forward* references to a behavior and makes self-recursive definitions possible. We use this particularly for defining `direction`.

---

[2] (_,_) is a Scala anonymous function that combines two arguments into a pair.

```
val svgFRP = new SvgFRP("playground", width, height)
val direction: DBehaviorᶜ[Vec2D] = {
  val previousPosition: Behaviorᶜ[Vec2D] = DBehaviorᶜ.delayed(position)
  val directionEv = previousPosition.snapshotWith(svgFRP.mousePosition) {
    (prevPos, mouse) ⇒ mouse - prevPos
  }
  directionEv.hold(Vec2D.zero)
}

val player: DBehaviorᶜ[Player] = -- (see before) --
val position: DBehaviorᶜ[Vec2D] = player.map(_.position)
```

The player always moves towards the user's mouse position. We define `previousPosition` by delaying the player's position with `DBehaviorᶜ.delayed`, obtaining the player's position just *before* the current. The current mouse position is retrieved through the `SvgFRP` object, which represents an SVG tag in the HTML interface and makes screen-to-SVG coordinate conversions. We define `directionEv` as the difference between the mouse position and the previous position. The final `direction` behavior is then defined, taking the initial direction as `Vec2D.zero`.

Whether a player is attacking is simpler: we simply look if the spacebar is down using the keyboard interface.

```
val kb = new Keyboard()
val attacking: DBehaviorᶜ[Boolean] = kb.isKeyDown(" ")
```

**The Game Interface**  A multi-tier FRP application is ultimately created by defining a value `ui` of type `DBehaviorᶜ[HTML]`. This behavior defines the value of the main HTML tag at every moment, and these values are rendered on screen. In this way, the programmer declaratively defines the application as "everything that is visible to the user". This main value is also a discrete behavior, i.e. it contains a notion of *initial* value and changes values at discrete times. Simply from this definition, the framework has all the information it needs to efficiently update the client's view.

In our example we want the interface to display the game. We obtain the SVG tag from the `SvgFRP` object by providing a camera and SVG tags. This camera is defined to show a fixed-size view of the game centered around the player. The behavior of SVG tags currently only contains the player's SVG representation. The resulting SVG tag is wrapped up in some HTML to produce our `ui`.

```
val camera: DBehaviorᶜ[Camera] = position.map { p ⇒
  Camera(p - Vec2D(cameraWidth / 2, cameraHeight / 2), cameraWidth, cameraHeight) }
val svgContent = player.map(p ⇒ List(p.svg))
val ui: DBehaviorᶜ[HTML] =
  svgFRP.svg(camera, svgContent).map { svg ⇒ section(article(svg)) }
```

Note that we use an HTML library that exposes HTML (and SVG) tags as regular Scala functions so that — as is common in multi-tier languages — interfaces are written with the full power of a general purpose programming language.

■ **Listing 3** Primitives for crossing from the Client to the Session tier and vice versa

```
object DBehaviorᶜ { def toSession[A](db: DBehaviorᶜ[A]): DBehaviorˢ[A] }
object DBehaviorˢ { def toClient[A](db: DBehaviorˢ[A]): DBehaviorᶜ[A] }
```

### 2.2 Multiplayer CircleRoyale

In a multiplayer version of CircleRoyale, we do not want to compute the state of the world locally on the client. Implementing this change requires surprisingly few changes. In Gavial, going from client to server is easy with tier-crossing primitives .toSession and .toClient as shown in listing 3.

The session tier is the server-side counterpart of the client tier we've been using so far: its behaviors and events live on the server side and there is one instance of it for each client. Crossing from the client to the session tier simply sends a client's value to that client's instance of the session tier, or vice versa. Using these primitives, we change the code as follows (unmodified code in yellow).

```
val sessionInterval = new ServerTick(1.second / 10).sessionElapsedTime
val serverInput: Eventˢ[(Vec2D, Boolean)] = Eventᶜ.toSession(throttledInput)
val playerInput: Eventˢ[(Boolean, Vec2D)] =
  serverInput.hold((Vec2D.zero, false)).sampledBy(sessionInterval)
val sessionPlayer: DBehaviorˢ[Player] =
  playerInput.fold(Player.default) { case (p, (dir, att)) => p.update(dir, att) }
val player: ClientDBehavior[Player] = DBehaviorˢ.toClient(sessionPlayer)
```

In this listing we do two things at once, (1) we do the previous computation (unmodified in yellow) at the *server* side on the *session* tier and (2) we make sure that the server computes *all* currently connected players (all instances of the session tier) at a rate of 10 Hz. The server updates at 10 Hz, even if 20 clients connect (which would produce input at 200 Hz).

To do achieve these two goals, we send player input to the server using toSession and re-throttle it to 10 Hz as before, using a server abstraction ServerTick. This rate drives the server computation and steps the game forward, sessionElapsedTime is the same across all sessions, as such everyone's game steps forwards at the same time. With *server-side* player input we define a sessionPlayer, with the same logic as before: user input updates the player's position and whether or not he is attacking. The client-side player definition that we used to define the interface is replaced by simply bringing the sessionPlayer to the client tier. These changes are all that is needed to construct a server-side version.

**Tiers: Client, Session and Application**    However, CircleRoyale is still not multiplayer. To add that functionality, we need a way to combine data from different clients. Gavial offers a third tier where this is possible: the application tier. The conversion functions to the application tier expose the session tier's multiplicity: a value in the session tier corresponds to a map of values, indexed by Client as shown in listing 4. Client values act as *connection* tokens: they identify a browser connection to the server. In

**Gavial: Programming the web with multi-tier FRP**

■ **Listing 4** Primitives for crossing from the Session to the Application tier and vice versa

```
object DBehavior^S { def toApp[A](db: DBehavior^S[A]): DBehavior^A[Map[Client, A]]
                     def client: DBehavior^S[Client] }
object DBehavior^A { def toSession[A](db: DBehavior^A[A]): DBehavior^S[A] }
```

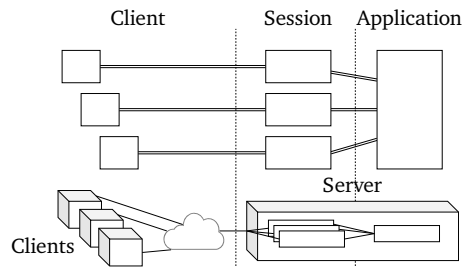

■ **Figure 3** The tiers available in Gavial.

the session tier, the `Client` identifier for the current connection is available through the `client` primitive.

A summary of the three tiers in Gavial is shown in figure 3.

**Application Tier and Client Tokens**   With the application tier, we can add user interaction to our game. We assume an implementation of a pure function `deadClients`, which performs a form of collision detection. It checks among all players that are still alive if one was hit by the weapon of another. We omit the implementation because it is not relevant to our discussion. The method returns all clients that have died (including those within the given `players` that were already dead).

```
def deadClients(players: Map[Client, Player]): Set[Client] = -- cut --
val playerInputAndDead: Event^S[(Boolean, (Vec2D, Boolean))] =
  DBehavior^S.delayed(dead).snapshotWith(playerInput) { (_, _) }
val checkedPlayer: DBehavior^S[Player] = playerInputAndDead.fold(Player.default) { (p, in) ⇒
  in match { case (dead, (dir, attacking)) ⇒ p.update(dir, attacking).setDead(dead) }
}
val checkedPlayers: DBehavior^A[Map[Client, Player]] = DBehavior^S.toApp(checkedPlayer)
val losers: DBehavior^A[Set[Client]] = checkedPlayers.map(deadClients)
val dead: DBehavior^S[Boolean] =
  DBehavior^A.toSession(losers).map2(DBehavior^S.client) {_ contains _}
```

Our previous definition of a player is no longer enough: we must now also track whether a player has died. We delay the (soon-to-be-defined) `dead` behavior and snapshot it with the `playerInput` from before and fold the result to compute a player including its `.dead` property. We then convert our `checkedPlayer` session value to an application value using `toApp`. The resulting behavior contains a map of all connected clients and their respective player states. The collision detection function `deadClients` is mapped over `checkedPlayers` to collect those who have lost the game, both old losers and new. With the set of dead clients, we define `dead` by checking whether or not the session tier's `client` (available through `DBehavior^S.client`) is among the dead players.

**Drawing Active Players**   At this point, the game is just missing an interface that shows all living players:

```
val survivors: DBehaviorᴬ[List[Player]] = checkedPlayers.map(_.values.toList.filter(!_.dead))
val svgContent: DBehaviorᶜ[List[HTML]] =
   DBehaviorˢ.toClient(DBehaviorᴬ.toSession(survivors)).map(_.svg)
val gameUI: DBehaviorᶜ[HTML] =
   svgFRP.svg(camera, svgContent).map2(DBehaviorˢ.toClient(dead)) { (svg, dead) ⇒
      section(article(if (!dead) svg else h1("You died!"))) }
val ui: DBehaviorᶜ[HTML] = gameUI
```

First, we filter out the survivors in `checkedPlayers` and send them to the session, and subsequently the client tier. The result is then turned into a discrete behavior of HTML tags and passed to `SvgFRP.svg`. Additionally, we send the existing `dead` to the client and use it to show the message "You died!" to dead players.

**Adding a Chat**   Gavial also supports "regular" HTML applications that are not so heavily SVG based. To demonstrate this, we extend CircleRoyale with a minimal chat, positioned underneath the game. Users can type in their name and message and all submissions are shown in a simple list. As soon as a user submits a message, his character is labeled so that people can identify messages' authors.

A chat message is represented as a `Message`, a Scala case class containing a name and message and a method for converting to a string representation.

```
case class Message(name: String, message: String) { val string = s"$name says $message" }

val msgSource: EventSourceᶜ[Message] = Eventᶜ.source[Message]
val msgs: Eventˢ[Message] = Eventᶜ.toSession(msgSource)
val appMsgs: Eventᴬ[Map[Client, Message]] = Eventˢ.toApp(msgs)
val chatInput: Eventᴬ[List[String]] = appMsgs.map(_.values.toList.map(_.string))

val chat: DBehaviorᴬ[List[String]] = chatInput.fold(List.empty[String]) { (acc, n) ⇒ n ++ acc }
val chatUI: DBehaviorᶜ[HTML] =
   DBehaviorˢ.toClient(DBehaviorᴬ.toSession(chat)).map { c ⇒ ul(c.map(msg ⇒ li(msg))) }
```

An event source is created, onto which messages can be pushed. Naturally, the chat is accumulated at the server side, by retrieving messages from the client and folding them into a list (most recent messages at the top). The accumulated `chat` is sent back to the client and rendered into an HTML list.

**Incremental Behaviors**   Unfortunately, the current chat implementation has a problem: how `chat` is sent to the client. This discrete behavior encodes *when* the list changes but not how. In other words, every time a new message is added to the chat, the chat log is seen as a completely new list of strings. When we send this behavior, this means that the full list will be transmitted to the client on every update, and network traffic will grow over time.

Behaviors are a more natural encoding of the chat log (as well as having other benefits such as automatic bootstrapping of clients, see section 3.5). So to solve this without using an encoding of the chat log through events, we add an additional FRP primitive: incremental behaviors [24]. They are behaviors that not only encode *when*

■ **Listing 5**  Incremental Behavior

```scala
trait Event[A] { def foldI(f: (B, A) ⇒ B): IBehavior[B, A] }
trait IBehavior[A, DA] { def toDBehavior: DBehavior[A] }
object IBehaviorᶜ { def toSession[A, DA](cb: IBehaviorᶜ[A, DA]): IBehaviorˢ[A, DA] }
```

■ **Listing 6**  DOM

```scala
object UI {
  def listen[R](a: Attr, src: EventSourceᶜ[R])(f: js.Dynamic ⇒ R): AttrPair[EventSourceᶜ[R]] }
object Eventᶜ { def source[A]: EventSourceᶜ[A] }
```

a value changes but also *why* it changes, they can be used to model incremental computations and can serve as a base for other work such as incremental collections [14]. We can also use them in Gavial to reduce the network payload.

As shown in listing 5, an incremental fold (`foldI`) on an event creates an incremental behavior. Incremental behaviors can be sent across all tiers, just like other FRP primitives. While they have their own (incremental) operations, for this example it suffices to know that they can be turned into discrete behaviors. We can now replace our suboptimal chat log implementation with a more efficient version, with minimal changes (marked in green, note the appearance of `I`s):

```scala
val chat: IBehaviorᴬ[List[String], List[String]] =
  chatInput.foldI((List.empty[String]) { (acc, n) ⇒ n ++ acc }
val chatUI: DBehaviorᶜ[HTML] =
  IBehaviorˢ.toClient(IBehaviorᴬ.toSession(chat)) .toDBehavior.map { c ⇒
    ul(c.map(msg ⇒ li(msg)))}
```

**Hooking into the DOM**  All that is left to complete the chat is the interface. This allows us to introduce Gavial's interface to HTML elements and their event handlers.

Listing 6 shows part of the API for DOM events. Event sources are events with an "open end" and with an imperative API through which non-FRP code can inject values. `UI.listen` takes an extra function which turns a dynamic Scala.js value into a value of type `R` and produces an attribute pair that can be used to install the appropriate event handler on an HTML tag, e.g., button(width := "5", UI.listen(onclick, src)(_ ⇒ 1)).

We redefine `ui` a final time. The main value of our application will now show both the game interface and the chat interface. Additionally, it contains a form with a submit button that is hooked up to the `msgSource` event source.

```scala
val ui = chatUI.map2(gameUI) { (chat, game) ⇒ div(game,
  form(input(`type` := "text", placeholder := "Name", name := "name"),
    input(`type` := "text", placeholder := "Message", name := "msg"),
    input(`type` := "submit"),
    UI.listen(onsubmit, msgSource) { ev ⇒
      val formElements = ev.target.elements
      val name = formElements.name.value.asInstanceOf[String]
      val message = formElements.msg.value.asInstanceOf[String]
      Message(name, message)}), chat)}
```

**Player Labels** Finally, we add player labels in game after they have posted to the chat. Although this code does not introduce new functionality of Gavial, it will allow us to explain an important aspect of tier-crossing (see section 3.6).

```scala
val optName: DBehavior^S[Option[String]] =
  msgs.map(msg ⇒ Some(msg.name)).hold(None).map2(dead) { (n, d) ⇒ if (!d) n else None }
val labelInfo: DBehavior^S[(Option[String], Vec2D)] =
  optName.map2(sessionPlayer) { (name, p) ⇒ (name, p.position)}
val allLabelInfo: DBehavior^S[List[(Option[String], Vec2D)]] =
  DBehavior^A.toSession(DBehavior^S.toApp(labelInfo).map(_.values.toList))
val clientLabels: DBehavior^C[List[Option[HTML]]] = DBehavior^S.toClient(allLabelInfo).map { ls ⇒
  import UI.html.{svgAttrs ⇒ a}
  ls.map { case (name, Vec2D(x, y)) ⇒ name.map { str ⇒ text(a.x := x, a.y := y, str) }
}

val svgContent: DBehavior^C[List[HTML]] =
  clientSurvivors.map2(clientLabels) { (ap, ls) ⇒ ap.map(_.svg) ++ ls.flatten }
```

This code collects the name and position on the session tier (for living players who have already posted a message), sends it to the application tier and back (to collect all labels), and next to the client, where the non-empty labels are added to the SVG element.

## 2.3 XHR or Websocket Backend

For implementing the client-server crossing primitives, Gavial can work in one of two ways: using XMLHttpRequests or using WebSockets. The former is more widely supported and does not require a long-running open connection for each client on the server, but the latter allows bidirectional communication.

CircleRoyale does in fact use bidirectional communication. Consider, for example, the definition of svgContent in the final multiplayer example. Remember that survivors is a discrete application-tier behavior of all living players in the game. It updates at a fixed rate of 10 Hz and its new value is then pushed to all clients. Because this rate is server-initiated, our implementation requires websockets.

Nevertheless, if we want to avoid web sockets, we can modify the game so that servers do not *push* values to clients, but clients *pull* from servers. We already have the client event time, a 10 Hz timer event. In polledPlayers, we re-use this time, send it to the server, turn it into a discrete behavior and use that behavior to read out values of survivors as polledPlayers. The polled players in turn get sent back to the client and define the new *xhr-compatible* clientSurvivors.

```scala
val svgContent: DBehavior^C[List[HTML]] = // unmodified
  clientSurvivors.map2(clientLabels) { (ap, ls) ⇒ ap.map(_.svg) ++ ls.flatten }
val polledPlayers: DBehavior^S[List[Player]] = // xhr-compatible
  DBehavior^A.toSession(survivors).sampledBy(Event^C.toSession(time).hold(0))
val clientSurvivors: DBehavior^C[List[Player]] = DBehavior^S.toClient(polledPlayers)
```

Note again that no other code needs to change. If we make similar changes for other server-initiated session behaviors that are sent to the client, our application becomes xhr compatible. In fact, since both client-to-server and server-to-client updates are both driven by the time event, messages to the server and responses to the client can be exchanged in a single HTTP request.

Programmers can force xhr-mode on expressions by placing asserts which will reliably get detected during development. We discuss both backends and their requirements of each in detail in section 3.7.

## 3  Making a Realistic Multi-tier FRP for the Web

After this hands-on introduction to Gavial we take a more detailed look at its main features. This includes both existing and novel ideas and shows that FRP applied to a multi-tier web setting can benefit the development of web applications.

### 3.1  Practical FRP and Incremental State

Since there exist quite a variety of FRP flavors in academic literature and in practical implementations, it is useful to take a moment to discuss where our API can be situated in the FRP family tree and which changes were made and why, to make it usable in practice.

We support both *discrete* and *non-discrete* behaviors. The latter are behaviors that may change continuously over time or at unknown times and we do not offer, for example, a method Behavior[A].changes for them. They are evaluated as needed, similar to [13]. This choice allows us to support behaviors that are not native to the FRP system for which changes are impossible or expensive to track, such as databases or DOM properties.

On the other hand, discrete behaviors additionally expose *when* a behavior changes value. An example where this is useful is the discrete client behavior ui of HTML tags. Because this is a discrete behavior, the programmer can define *when* the DOM should be updated. Discrete behaviors can be converted to general behaviors, simply by throwing away the "when" information.

Finally, as explained before, we also make use of incremental behaviors, which reify the fold operation on events [24] and expose not only *when* (like discrete behaviors) but also *how* a behavior changes its value. As shown in the CircleRoyale example, incremental behaviors allow us to implement efficient tier-crossing primitives without forcing programmers to use an unnatural representation of behaviors. They are created by folding events and expose both changes and deltas, the change to the behavior as a result from f on the event from which it was folded, and the value that initiated this change deltas.

```
trait Event[A] { def foldI(init: B)(f: (B, A) ⇒ B): IBehavior[B, A] }
trait IBehavior[A, DA] { val initial: A
                         val f: (A, DA) ⇒ A
                         def changes: Event[A]
                         def deltas: Event[DA] }
```

They can be converted to discrete behaviors by dropping the *why* information of deltas. However, discrete behaviors can also be treated as a special case of incremental behaviors. In this case, the `changes` match the `deltas` and the folding function `f` simply ignores the older value while using the new value: `(_, a) ⇒ a`. We use this property frequently to re-use incremental behavior specific APIs.

**First-order FRP**   We also limit our FRP to *first-order FRP* (as opposed to higher-order FRP). In other words, we do not offer APIs like `flatten: Behavior[Behavior[A]] ⇒ Behavior[A]` that flatten nested FRP abstractions. First-order FRP is conceptually simpler, because dependencies between behaviors and events are statically known. These guarantees make it suitable for multi-tier FRP, as dynamically generated client/server crossing would be hard to understand and implement. It also avoids certain tricky problems of higher-order FRP, like the so-called *time leaks* that cause memory leaks in naive higher-order APIs (see, for example [32]), and we do not need to modify our API to prevent them. On the downside, first-order FRP is less expressive, but as shown in [36], the full expressiveness of higher order FRP is not always necessary.

### 3.2 Tiers

Our API is tailored to the standard web distribution model, where there is essentially one server and an arbitrary number of active clients (browsers) that connect to the server. We assume that these clients are only active for a subset of the application's lifetime and we distinguish the programs in multiple tiers. Previous work in multi-tier languages in general [8, 19, 30, 6, 1] or specifically in multi-tier FRP [25] work with two tiers: client and server. A problem with a two-tiered system is that the framework makes a decision to focus itself to one style of programs. Regular request-response interaction between the client and the server is easier if the chosen server tier is most akin to the session tier. Typical create-read-update-delete applications fall in this category. On the other hand, applications that rely heavily on user-to-user interaction through the server are more difficult to write and have to imperatively manage state across clients somehow. Such interactive applications are easier to write if the server tier is most akin to the application tier since sharing state across clients is part of the programming model. However, programs that primarily focus on handling a single client's requests become very tedious to write.

As such, we support both types of programs in an equally convenient way through the three tiers previously explained and illustrated in figure 3: the application (single instance, server-side), session (client-specific, server-side) and the client tier (client-specific, client-side). The server keeps track of every active client connection and assigns it a unique identifier. This value is exposed in the API as opaque values of type `Client` and shows up in certain tier-crossing primitives (e.g., `.toApp`) as well as in the following primitives:

```
object BehaviorS { val client: BehaviorS[Client] }
object EventA { val clientChanges: EventA[ClientChange] }
object IBehaviorA { val clients: IBehaviorA[Set[Client], ClientChange] }
```

The `client` primitive exposes a session's `Client` as a session behavior. The application event `clientChanges` informs about clients connecting or disconnecting. We use Scala's sealed traits to encode the event information (the `Client` and whether it just connected or disconnected):

```scala
sealed trait ClientChange { val client: Client }
case class Connected(client: Client) extends ClientChange
case class Disconnected(client: Client) extends ClientChange
```

### 3.3 Crossing Application & Session Tier

When sending events or behaviors between the session and application tier, the primitives need to deal with the fact that the session tier exists in many copies at the same time (one for each active client), while there is only one instance of the application tier.

This is reflected in the type of the session/application tier-crossing primitives for *events*:

```scala
object Event^S[A] { def toApp(e: Event^S[A]): Event^A[Map[Client, A]] }
object Event^A[A] { def toSession(e: Event^A[A]): Event^S[Map[Client, A]] }
```

Sending a session event to the application tier produces an `ApplicationEvent` for a different type of values: `Map[Client,A]` instead of `A`. Intuitively, the event `.toApp(e)` will fire whenever at least one of the copies of the session event `e` fires and a map will be produced containing the identifier of the connection and the event value for each of these copies. Conversely, an application event `e` can be sent to a session event of type `Event^S[A]` where it fires for each client whenever `e` fires.

The situation is similar for sending *behaviors*.

```scala
object Behavior^S[A] { def toApp(b: Behavior^S[A]): Behavior^A[Map[Client, A]] }
object Behavior^A[A] { def toSession(b: Behavior^A[A]): Behavior^S[A] }
```

Sending a session behavior to the application tier creates a `Behavior^A[Map[Client,A]]` which maps active clients to their value of the behavior. The converse primitive simply produces a behavior with the same value for every client.

**Incremental & Discrete Behaviors**  Sending incremental behaviors between the session and application tiers is a bit more complicated. We do not discuss discrete behaviors, they can be seen as a special case of incremental behaviors and all techniques discussed here are valid for those.

Consider first a session incremental behavior `b` and think about what the type of `IBehavior^S.toApp(b)` should be if `b` has type of values of `A` and type of deltas `DA`. As before for regular behaviors, the type of values for `Behavior^S.toApp(b)` should naturally be `Map[Client,A]`: a map containing for every active client the value of the corresponding copy of the session incremental behavior. But now we should choose a type of deltas that can represent any way in which the value initiates change. Obviously, one or more of the copies of `b` may change, so we need this type to contain a `Map[Client, DA]`. However, another possible cause for the value to change is that clients have connected to the server, in which case a new copy of the session behavior will be made and a new entry in the map will appear, and conversely clients may disconnect in which case a copy will be dropped and an entry will disappear. This is why the type of deltas should be (`Map[Client, DA]`, `Option[ClientChange]`), that is, each delta is a map of client specific changes (possibly empty) and a possible change in client connections. Both deltas can also appear at the same time, for example, if the value delta is derived straight from the `clientChanges` primitive. The complete type of the method:

```
object IBehaviorˢ { def toApp[A, DeltaA](sb: IBehaviorˢ[A, DeltaA]): IBehaviorᴬ[Map[Client, A], (
    ↪ Map[Client, DeltaA], Option[ClientChange])] }
```

The other way around, sending an application incremental behavior to the session tier simply sends values and deltas directly:

```
object IBehaviorᴬ { def toSession[A, DeltaA](sb: IBehaviorᴬ[A, DeltaA]): IBehaviorˢ[A, DeltaA] }
```

Note that crossing from the session to the application tier is done in-memory. As such, bandwidth overhead is less of an issue and we expect the incremental session to application API to only be used in applications where incremental behaviors are used to also reduce *computational* overhead. If computational overhead is not a main issue, incremental behaviors can be turned into discrete behaviors with major simplifications in the API (identical to `Behaviorˢ`).

### 3.4 Crossing Client & Session Tier

Being able to send events and behaviors from a client tier to a server tier is one of the key features of our model. We do not offer tier-crossing primitives for non-discrete behaviors between the client and session tier. The reason is that we want to use only asynchronous communication between client and server. Imagine a tier-crossing primitive for a client (non-discrete) behavior `b` to the server side as `b.toSession`. It is generally impossible to predict upfront at the client side when the value of `b.toSession` will be required on the server, so the only possible implementation would have the server synchronously request the current value to the client and block execution until the client answers.

The client/session tier-crossing primitives need to transmit values across the network. This requirement shows up in the API as type-classes encoded as Scala's implicit arguments [21, 20]. All values that cross the network are required to be serializable, visible in our API as extra requirements on the type in form of: `A: Encoder: Decoder`. We use an existing Scala library to supply encoders and decoders for standard items and for case classes (semi-)automatic derivation is available.

Between the client and session tier, sending *events* is straightforward. A client event `e` of type `Eventᶜ[A]` can be sent to the server as `.toSession(e)` of type `Eventˢ[A]` and vice versa using `.toClient(e)`. Intuitively, when the event `e` fires, it asynchronously sends to the other end of the tier boundary. At that other end, the event fires from the sent event after the network delay when its received.

```
object Eventᶜ { def toSession[A: Decoder: Encoder](e: Eventᶜ[A]): Eventˢ[A] }
object Eventˢ { def toClient[A: Decoder: Encoder](e: Eventˢ[A]): Eventᶜ[A] }
```

For an *incremental behavior* `b`, it is known when the value changes, so we can produce correct values at the other side of the network for the sent behavior `.toSession(b)` or

.toClient(b) simply by sending an update whenever b changes. As explained before, this does not require transmitting the full value of behavior b, but just the deltas that represent what has changed. We can re-compute the new value on the server from the previous value and the delta. In other words, if b has type IBehavior$^c$[A,DA], we only need to transmit the value of type DA when b changes and the result is an incremental behavior of type IBehavior$^s$[A,DA]. Of course, when the client first connects, there is no point in transmitting a delta and we transmit the full initial value.

```
object IBehaviorˢ { def toClient[A: Decoder: Encoder, DeltaA: Decoder: Encoder](b: IBehaviorˢ[A,
    ↪ DeltaA]): IBehaviorᶜ[A, DeltaA] }
object IBehaviorᶜ { def toSession[A: Decoder: Encoder, DeltaA: Decoder: Encoder](b: IBehaviorᶜ[A,
    ↪ DeltaA]): IBehaviorˢ[A, DeltaA] }
```

### 3.5 Bootstrapping Clients

One of the useful properties of combining FRP with multi-tier languages is automatic bootstrapping of clients [25]. Bootstrapping is the initial provisioning of client values with the latest state of session behaviors sent to the client. It is a standard task in web application development, typically solved in an application-specific way by for example, embedding initial values in the HTML or by polling for the latest values at client startup. The multi-tier FRP abstractions of Gavial allow us to solve the bootstrapping in a general, natural and transparent way.

This property is a direct result of the (natural) semantics of .toSession on incremental behaviors, which define the initial value of an incremental session application behavior b as the value of b at the connection time of the client. If an application behavior sent to the session tier is further sent to the client, the client will also be provisioned with this value. This saves developers the work of implementing manual initialization schemes and automatically helps them define the initial state of new clients.

### 3.6 Tiered Glitch Freedom with Minimal Overhead

Something that we have not discussed before are the guarantees of the tier-crossing primitives and how they differ from regular FRP semantics. Correctly implemented FRP libraries follow FRP semantics and protect programmers from partial event propagation. For example, you would expect t in the following expression to remain true throughout updates to x:

```
val x: Behavior[Int] = -- cut --
val y: Behavior[Int] = x.map(_ + 1)
val t: Behavior[Int] = x.map2(y)(_ < _) // true
```

Propagating x = 20 from x = 1 should evaluate 20 < 21 for b instead of ever ending up in 20 < 2 or 1 < 21. If such partial updates can be observed, they are called *glitches*. Our proposed multi-tier FRP has a similar property for network-crossing primitives. To explain this, we visualize propagation of the above toy example as a graph in figure 4 and add a network between x and y, and t. This corresponds to the code in the left part of figure 4

```
val x: DBehaviorᶜ[Int] = -- cut --
val y: DBehaviorᶜ[Int] = x.map(_ + 1)
val t: DBehaviorˢ[Int] =
    DBehaviorᶜ.toSession(x).map2(DBehaviorᶜ.toSession(y))(_ < _)
```

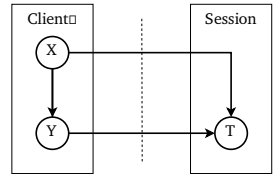

■ **Figure 4**   x < y across tiers

```
val x: DBehaviorᶜ[Int] = -- cut --
val y: DBehaviorᶜ[Int] = x.map(_ + 1)
val t: DBehaviorᶜ[Int] =
    DBehaviorˢ.toClient(DBehaviorᶜ.toSession(x)).map2(y)(_ < _)
```

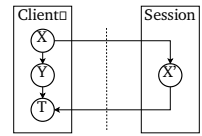

■ **Figure 5**   x ≮ y across tiers

In a naive implementation of multi-tier FRP (such as [25, 23, 4]) glitches would inevitably occur in t due to network delays, however in our proposal neither client to server communication nor server to client communication results in glitches.[3] In Gavial, all events or behaviors that cross from the client to the session tier and vice versa are propagated atomically. This means that t will always be true, no partial updates can ever reach the session tier.

Glitches are still possible, but only if FRP values depend on values of different tiers, for example as follows:

In this case, there will be a delay between the updates of y and the version of x sent to the server and back. We purposefully do nothing to hide such network delays, as that would require the computation on the client (or server if the situation was reversed) to be blocked until the full propagation catches up. We leave it up to the programmer to take network delays into account. In other words, we prevent those forms of glitches that can be prevented without working around the distribution model of the web. We will refer to this property as "Tiered Glitch Freedom". It is not the purpose of this paper to express or prove this property formally, but it is in fact expressed by our denotational semantics (see section 3.9).

Perhaps surprisingly, we rely on this property in two cases of the CircleRoyale example. Once in the interaction of defining `direction` and showing all players on the user interface and another time in the interaction between naming players and showing them. In both of these cases we are sure that: (1) a player's direction is calculated based on the position that is actually shown in the playing field and (2) a name in the chat is always drawn at the same time on the playing field. There can be no consistency mismatches between names, direction or what is visible in the playing field even though we conceptually do three separate tier-crossings: `sessionPlayer` (for direction), `clientSurvivors` (for drawing players) and `name` (for labeling drawings). Our tier-crossing primitives guarantee that all values cross the network atomically within one propagation cycle. This property allows programmers to more safely refactor and add to existing multi-tier FRP code with the added guarantee that data does not get propagated in unexpected ways. With a naive multi-tier FRP implementation we would have had to go back and manually batch updates into a single .toClient call. This same property is available in the other direction as well, all toSession crossings are done in the same atomic manner.

---

[3] With the exception that it is possible to make use of the FFI or Async to intentionally separate network propagation for efficiency.

Tiered glitch freedom gives us strong guarantees: manipulating behaviors or events on one tier and then sending to another is equivalent to first sending and then manipulating them. For example, the previous definition of t in figure 4 is (functionally) identical to the following definition:

```
val t: DBehaviorˢ[Int] = DBehaviorᶜ.toSession(x.map2(y)(_ < _)) // true
```

Our approach avoids a large implementation cost (as seen in section 4) but still provides useful guarantees. It forms a middle ground between two extremes "local glitch freedom" and "total glitch freedom" that we call *tiered glitch freedom*.

**Local Glitch Freedom**   Client/server web applications are usually treated as separate programs and previous work on multi-tier FRP also treats client and server programs essentially as distinct but connected FRP applications [25, 23, 4]. Instead of having consistency guarantees regarding glitches, cross-tier connections are treated as a communication channel that lies beyond the scope of the FRP semantics. In practice this means that events are transmitted as soon as they come in but the guarantees we have for t in figure 4 do not hold.

**Total Glitch Freedom**   The other extreme alternative to our tier-crossing semantics is glitch-freedom in a distributed setting [11, 16, 18]. In other words, across the distributed reactive program there can never be an observable partial propagation. To make this possible there are extra restraints on the propagation, certain projects require a global lock on the whole program, others block and queue propagation in certain subgraphs. In a web settings, for example if input validation is done on the server, then these semantics require that input from the client is transmitted to the server, validated and transmitted back, all in the same propagation, so that the client and server both block until everything is finished. To combat these issues, projects such as DREAM [16] and ScalaLoci [35] make these algorithms pluggable, but none have tiered glitch freedom as an option.

### 3.7  XHR or WebSockets?

A novel aspect of Gavial is that it automatically selects the network communication backend to use based on the primitives used to write the program. The xhr-mode can be used as long as the application does not require the server to initiate communication with a client. WebSocket mode becomes a requirement as soon as functionality cannot be implemented in a request-response style manner. Intuitively this happens in two cases: (1) whenever a server sends something to a client on its own (through timers

■ **Listing 7** Imperative FRP API

```scala
object Event { def source[A]: EventSource[A]
              def sourceWithEngineEffect[A](eff: (A ⇒ Unit) ⇒ Unit): EventSource[A] }
object Behavior { def sink[A](default: A): BehaviorSink[A]
              def fromPoll[A](f: () ⇒ A): Behavior[A] }
trait Engine { def fire(pulses: Seq[(EventSource[A], A) forSome { type A }]): FireResult }
```

or through the foreign function interface) or (2) whenever clients send information to other clients through the server.

To decide whether such cases are present, Gavial analyses the FRP graph and tags every event or behavior as "needing bidirectional communication" or not. Operations such as map simply take the mode of the parent event or behavior. Operations that combine multiple events or behaviors such as map2 take the most restrictive mode of its dependencies, if one requires bidirectional communication then so does the result.

An exception to this rule are the snapshotWith operations. Snapshotting a behavior b that requires bidirectional communication with an event e that does not, produces a result that does not require bidirectional communication either. This makes sense because changes in b will not cause changes in b.snapshotWith(e).

Calculating whether or not the bidirectional communication is necessary is done at startup time and developers can place asserts to force xhr-mode. The same functionality can be modeled in the Scala type-system, however, this would occur everywhere in the API and would clutter it significantly. We opted for a run-time implementation that does not use the type-system, however xhr-mode violations are reliably detected during development and stops execution with the appropriate error message before the web-server becomes available.

### 3.8 Interacting with the World

We have seen a glimpse of how to interact with the world in section 2. However, most of the actual interaction was hidden behind some convenient abstractions such as SvgFRP. While these foreign APIs are not part of the core design effort, they make Gavial realistic and practical.

**Connecting to non-FRP APIs**   There are three main ways of interacting with non-FRP APIs from within the FRP system, through event sources, by polling behaviors and through behavior sinks as shown in listing 7.

Event sources are "open" events. They have the added ability of being triggered imperatively through an "engine". The engine is an exposed value of the underlying FRP library and contains a fire method that starts a propagation cycle in the FRP network. Another way of making an event source is through the sourceWithEngineEffect method, this requires a function that gets a function as a parameter of type A ⇒ Unit. The given function imperatively fires a value onto the event source that is created through the method and allows programmers to conveniently write code that interacts with the DOM, for example an excerpt of the Keyboard class:

**Gavial: Programming the web with multi-tier FRP**

◼ **Listing 8**   DOM

```
object UI {
  def listen[R](a: Attr, src: EventSourceᶜ[R])(f: js.Dynamic ⇒ R): AttrPair[EventSourceᶜ[R]]
  def read[R](tag: HTML)(sink: BehaviorSinkᶜ[R], selector: js.Dynamic ⇒ R): HTML }
```

```
def keyEvSrc(name: String): EventSourceᶜ[Key] =
  Eventᶜ.sourceWithEngineEffect[Key] { (fire: Key ⇒ Unit) ⇒
    @client val _ =
      dom.window.addEventListener[KBEvent](name, ev ⇒ if (!ev.repeat) fire(ev.key)) }
```

In this case the Scala.js DOM APIs are used to attach an event handler to the top window object. The handler uses the fire function to send keypresses directly to the event source that is being made. @client is an annotation that is required to use Scala.js specific APIs in the multi-tier section of a program, more about this in section 4.1.

For behaviors there are two options to interact with the outside world: polling behaviors and behavior sinks. A behavior created through fromPoll creates thunks around a function. A thunk is created on every propagation cycle and is forced whenever a value from the behavior is required, inside *one* propagation cycle a fromPoll behavior always returns the same value. A behavior sink is very similar except that it makes the polling function re settable. As long as no poll-function is set it returns the supplied default, it is heavily used to add property support in the DOM API [25].

**Builtin DOM Support**   The DOM API incorporates techniques discussed previously by Reynders, Devriese, and Piessens [25]. We give a brief overview, for a more detailed explanation on the DOM API and its design decisions we refer to that work. In the CircleRoyale example only half of the UI API is used, the full API supports listening to DOM events and reading from DOM properties as shown in listing 8.

Listening to events in the DOM is done by creating additional attributes with listen. These special attributes are created with a function and an event source where the given function (f) takes a dynamic Scala.js value and transforms the DOM event to a concrete result that has to match the type of the given event source. DOM Events are propagated to the FRP program as long as the special attribute is attached. Not just events are supported, by placing behavior sinks on an HTML tag it is also possible to read from DOM properties. The selector function is used to read from the element into the sink and similar to listen, properties are read as long as the special tag is in use.

**Asynchronous FRP**   For now, our FRP system executes single-threaded. However, we support Elm's asynchronous FRP [9] which allows the programmer to break out of ordered event processing and enable concurrent execution within FRP programs.

```
object Async { def execute[A](ev: Event[IO[A]]): Event[A] }
```

Through Async it is possible to execute an IO[A] and retrieve an A in a different propagation cycle on the resulting event. We use a library implementation of the IO monad for Scala which has both a JavaScript and a JVM implementation and allows developers to communicate asynchronously with external services such as other web APIs or databases without blocking the FRP program.

### 3.9 Denotational Semantics

While we hope that the API is intuitive and easy to comprehend, it is of course important to specify the semantics of the API completely and precisely. We defined a denotational semantics for Gavial as a non-ambiguous reference specification of the core APIs. Time and network delays were modeled and the semantics were actively used during API design. They helped us get the types of the tier-crossing APIs right and gave us useful insight when dealing with corner cases (particularly related to bootstrapping, see section 3.5). The denotational semantics do not play a large part in this paper and we do not use them to prove novel properties, but they were helpful as an implementation specification and might be helpful as a reference to the reader, it is available as supplemental material on https://doi.org/10.5281/zenodo.3647731.

## 4  Implementation

### 4.1 Embedded as a Library in Scala

Gavial is completely embedded in Scala in order to use existing libraries and it makes use of Scala.js [10], a Scala to JavaScript compiler. It is set up using the same techniques used in the Scalagna project [27], an experimental multi-tier-as-a-library for Scala. Gavial is implemented as two Scala libraries: a JVM and JavaScript library, as well as some common shared code. The server and client-side FRP primitives are respectively backed by real and mock implementations on the server and vice versa on the client and we make sure tier-crossing primitives are supported appropriately on each side. For more detail on how we use both compilers we refer to appendix A.

From a Gavial user's perspective, a program is a cross-build between two environments — the JVM and JavaScript backends — which is supported in the Scala build tool (SBT) using plugins. Since Gavial is just a library this means that a developer gets nice integration into known production quality Scala tools in comparison to creating a new multi-tier language from scratch.

**Reusing the Scala/Scala.js Ecosystem**  Developers can also make use of the entire Scala/Scala.js ecosystem of libraries as well as the JVM and JavaScript ecosystems through the corresponding Scala FFIs. Libraries that are only supported for one backend can be integrated, either in a backend-specific source file, or by using special (and somewhat crude) annotations @client and @server. @client-annotated code is not compiled on the server and vice versa.

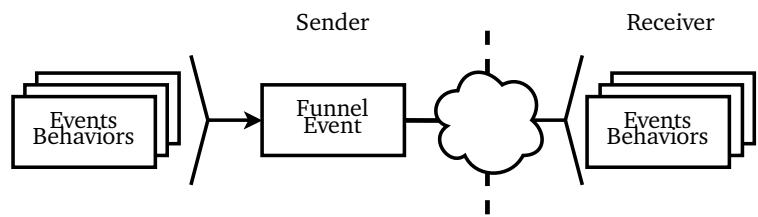

**■ Figure 6** Efficient Tiered Glitch Freedom

## 4.2 Efficient Tiered Glitch Freedom

As we have seen, client-server communication should behave with minimal glitches. To achieve this, all sent behaviors and events have a unique identifier, which is the same in the client and server code. To cross tiers with tiered glitch freedom we merge all events and behaviors into one large funnel event on the sender side. On the server-side, it is of type `Client ⇒ List[Message]` and contains a function that for all clients produces an (optionally empty) list of messages. Messages consist of the tier-crossed event or behavior's identifier and a value. At the receiver's side, a message router receives this list of messages, splits it up into updates that can be fed into the local FRP engine in a single propagation cycle. This whole process is shown in figure 6. This shows that implementing tiered glitch freedom is low-cost both in implementation as well as in performance cost in an FRP program.

**Backends**   The exact implementation depends on the backend the program is running on. The websocket backend simply uses bidirectional communication as expected. Whenever a propagation cycle ends on one end the changes are sent to the other and vice versa. The xhr backend works a bit different since client propagation cycles take care of both directions of communication. At any time the client propagation cycle requires events to be propagated, a request is sent to the server. The server running in xhr-mode then executes a propagation cycle which creates new values that should be shipped back to the client.

**Performance**   A small test in our student lab on CircleRoyale (the full version on websockets available on github.com/tzbob/circleroyale) was run on one PC while continuously adding players to the game. On an unoptimised single-threaded version of CircleRoyale, we were able to sustain 35 concurrent clients, i.e. ≈350 client-to-server messages and ≈2100 server-to-client messages per second. The most taxing operations were in collision detection (naive implementation) and the underlying messaging library. This limited test indicates that Gavial does not impose a large overhead compared to the underlying tried-and-tested Scala libraries.

## 4.3 Crossing Tiers with Incremental Behaviors

Propagating changes of behaviors is entirely similar to those of events. However, extra support is needed to replicate initial values. Similarly to merging all changes into one big event, behaviors are all bundled into a single behavior of type `Client ⇒ List[Message]`.

■ **Table 1** Comparison Table for Multi-tier Reactivity: ✓→ has feature; X→ does not have feature; empty → not applicable; other → in table

| Project | Reactive | RP Tier-Crossing | Consistency | Overhead | Incremental | Web API | Base Language |
|---|---|---|---|---|---|---|---|
| Gavial | Both | ✓ | Tiered | ✓ | ✓ | ✓ | Scala |
| Ur/Web | Client | To Client | Tiered | ✓ | X | ✓ | From scratch |
| ScalaLoci | Both | ✓ | Flexible | Flexible | X | | Scala |
| DREAM | Both | ✓ | Flexible | Flexible | X | | |
| QPROPd | Both | ✓ | Total | X | X | | |
| SID-UP | Both | ✓ | Total | X | X | | |
| Eliom | Both | ✓ | Local | ✓ | X | ✓ | OCaml |
| Scala Multi-tier FRP | Both | ✓ | Local | ✓ | X | ✓ | Scala |
| AmbientTalk/R | Both | ✓ | Local | ✓ | X | | From scratch |
| Flask | Both | ✓ | Local | ✓ | X | | Haskell |
| (Hip)Hop | Both | ✓ | | | | ✓ | Scheme |
| (Hip)HopJS | Both | ✓ | | | | ✓ | JavaScript |

The data to properly initiate the incremental behaviors is sent when a client connects to the server. Exactly when and how depends on the backend. In websocket mode the initial data is pushed from the server to the client as soon as a connection is made. In xhr-mode a request is sent from the client as soon as the client-side program is loaded.

## 5 Related Work

In this section we discuss distributed reactive and/or multi-tier programming languages and relate them to our work. We do not go into detail on multi-tier language proposals that do not have reactive programming features such as the initial proposed multi-tier calculus or later additions [19, 7] nor ML5 [34], Links [8] or automatic slicing techniques such as Stip.JS [22].

Regarding multi-tier languages, we look at languages that are based on existing languages such as Eliom [23] (OCaml), ScalaLoci [35] (Scala), Hop [29, 31] (Scheme), HopJS [30, 33] (JavaScript) and Flask [15] (Haskell) as well as languages that are built from scratch such as Ur/Web [6, 5] and AmbientTalk/R [4].

In the field of distributed reactive programming there are several programming languages or even algorithms that describe systems relevant to our multi-tier implementation of FRP with tiered glitch freedom such as SID-UP [11], DREAM [16] and QPROPd [18].

We focus on whether or not there is support for reactive programming on the client, the server, both or on flexible tiers[4] (also written as both). For these flexible projects we specifically look at the availability of reactive tier-crossing primitives and the consistency properties thereof.

---

[4] Some multi-tier languages support distributions other than the client/server architecture of the web.

### 5.1 Local Reactive Programming in Multi-tier Languages

Both Hop (based on Scheme) and its successor HopJS (based on JavaScript) have reactive programming libraries named HipHop(JS) [3, 33] respectively. The HipHop libraries are based on synchronous programming languages such as Esterel by [2] and make it possible to create reactive programs in a synchronous DSL similar to Esterel. Synchronous programs are written in isolation and plug into the regular Hop execution as input to output event processors. HipHop supports both execution on client and server-side of Hop but does not provide any means to create one conceptual reactive program across tiers and thus does not provide a means for automatic bootstrapping nor any cross-tier reactive consistency guarantees.

Ur/Web provides a *source* that can be compared to an *EventSource* that we discussed throughout the dissertation. It has the same imperative functionality as references, you can create them and set or get its value. Only creation or setting the source is supported on the server. Composing sources is done by "subscribing" to a source and creating a *signal*. Such a signal allows composable reads over several sources and can be embedded into Ur's HTML pages. This gives developers an imperative RPC-style interface from the server to a client-side FRP program. The entire page is created from the current source values and as such Ur/Web has a similar elegant solution to the bootstrapping problem we describe in section 3.5.

### 5.2 Multi-tier Reactivity

Embedding reactive programming in multi-tier programming by making it possible to write reactive programs in each tier is a first step. Several languages go further (like we do) and allow building a reactive program that spans all tiers with primitives to cross tiers. We divide related work in three sections of multi-tier reactivity, those that provide local glitch freedom, total glitch freedom and tiered glitch freedom.

**Local Glitch Freedom**    The simplest form of tier-crossing primitives provide local glitch freedom.

Multi-tier FRP in Scala [25] provides .to(Client/Server) on events and discrete behaviors, but naively connect client FRP applications to server FRP applications without minimizing glitches or providing any consistency guarantees. They solve the bootstrapping problem as we do in section 3.5.

Eliom provides a client/server reactive abstraction (since v5.0).[5] They provide a client *signal* that can be initialized on the server and used on the client. They also provide a server *signal* and have similar tier-crossing primitives that naively propagate events from one tier to the other. As such, they provide a similar solution to the bootstrapping problem as well as a multi-tier reactive programming environment similar to [25].

---

[5] https://opam.ocaml.org/packages/eliom/eliom.5.0.0/

An extension of AmbientTalk/R to combine the advantages of loosely-coupled publish/subscriber systems with the elegance of reactive programming constructs is explained in *Loosely-Coupled Distributed Reactive Programming in Mobile Ad Hoc Networks* by [4]. They provide *ambient behaviors* which is a construct that allows the propagation of events to reactive values hosted on other reactive networks by means of publish/subscribe. An *ambient behavior* is a behavior that is subscribed to previously exported behaviors. Our approach can be compared to theirs by looking at to(Client/Server) as a combination of export/subscribe. Since we assume a 'single server with multiple clients' architecture we greatly simplify our API, as a result we do not provide the flexibility that AmbientTalk/R provides.

Flask is not a multi-tier language applied to the client/server web, it is a distributed FRP language for sensor networks. They provide support for broadcast topologies and have no consistency guarantees regarding propagation.

**Total Glitch Freedom** Other than specifically targeting web development, academia also focused on a more general *distributed reactive programming* (DRP) with the aim of providing alternatives to the Observer pattern in a distributed environment. An overview of requirements and challenges of DRP is provided in *Towards Distributed Reactive Programming* by [28]. The projects we compare with in this space are not multi-tier languages specifically targeted toward the client/server nature of the web. They are targeted towards a larger distribution pattern of a reactive program where, e.g., multiple distributed reactive expressions make up a single program. In our distributed multi-tier project it is about how to unify a client and a server reactive program. Nonetheless, the programming models they build and propose are very related to our multi-tier reactive programming.

A DRP approach that focuses strongly on consistency guarantees is defined in [16]. They deliver three levels of consistency guarantees: causal, glitch free and atomic. Causal consistency refers to propagation that maintains causality within one process, e.g., $e_1$ happens before $e_2$ in the origin reactive nodes and will only be able to be observed in that order by other reactive nodes. Glitch free consistency means that a *partially* propagated FRP network is never observable, even in the distributed setting. Finally, atomic is a consistency guarantee that delivers *total* FIFO ordering and glitch freedom and thus is the most expensive of them all. Their implementation for glitch free consistency (including atomic, which adds distributed locking to it) requires cross tier propagation messages to include extra details (the history of the propagation) which causes the network traffic to increase. While their consistency guarantees are flexible, they do not provide a consistency guarantee that is similar to tiered glitch freedom.

Several other distributed reactive algorithms were proposed with similar goals. SID-UP [11], a distributed glitch-free propagation algorithm that minimizes messages compared to DREAM and requires a centralized "lock" that makes the distributed program unable to process more than one propagation at a time. QPROP [18], an algorithm that provides distributed glitch-free propagation that does not require a central coordinator for locking.

ScalaLoci is not a multi-tier programming language applied to the web, however, it is very related to our work. They also target the Scala language and also do this without modifying the compiler. Instead of having two (or three) set tiers they provide a type system in which a programmer can express the distribution of the program. The placement types are used to define on which location certain expressions live and they support reactive programming with tier-crossing primitives. The consistency guarantees of these tier-crossing primitives are flexible and pluggable, so far they support SID-UP and a propagation similar to [17] which provides no distributed guarantees. We think a version of tiered glitch freedom would complement the project well.

**Tiered Glitch Freedom**   Ur/Web provides client-side reactive programming but also has an interesting consistency property for its server-to-client tier-crossing primitives. There are no formal semantics on the Ur/Web RPC calls but if we understand correctly, Ur/Web provides a consistency property similar to tiered glitch freedom in the direction of server-to-client. Ur/Web's programming model is tied tightly to the request-response style of the web, all server-to-client communication within a response of a client-to-server RPC call is done atomically.

In comparison, our work has the same consistency guarantee, but instead of only providing it in one direction, our tiered glitch freedom (see section 3.6) has the same guarantee in both directions.

**Incremental Propagation**   None of the multi-tier reactive languages and algorithms we described document support for incremental propagation. They have no primitives similar to our `.toClient/Session` for incremental behaviors. As such, incrementally built behaviors such as an incremental collection propagate their full state instead of their change.

**Three Tiers**   Our multi-tier reactive programming model is split into three tiers. As earlier explained in section 3.2, the motivating factor is to cope with both request-response style applications as well as user-to-user style applications. In ScalaLoci it is possible to express "tiers" in the program manually, for example our client-session-application tiered system:

```
@peer type Client <: { type Tie <: Single[Session] }
@peer type Session <: { type Tie <: Single[Client] with Single[Application] }
@peer type Application <: { type Tie <: Multiple[Session] }
```

The APIs they provide have similar types as ours, that is, crossing from a single tie to a multiple tie shows the multiplicity in the primitive's type. `Signal[T].asLocalFromAll` on a session signal would return a `Signal[Map[Remote[Session], Signal[T]]]`. Glitch-freedom between the session and application tier for our project is a given since they run in the same FRP application on the server. As far as we know, ScalaLoci does not take into account deployments that are physically on the same JVM, so a session to application model we describe would incur the same overhead or glitches as the client to session tier connection would.

## 6 Conclusion and Future Work

In this paper, we have focused on the idea of multi-tier FRP (specifically for the web's client-server architecture) with asynchronous tier-crossing primitives. Several existing and novel ideas in both FRP and multi-tier research fit together to form Gavial. The core API and primitives (FRP with crossable tiers) were enriched with asynchronous behaviors, APIs to work with imperative programs, HTML support, etc., to support real-world applications. Novel ideas such as a three-tier model, minimal glitches and support for XHRs are available as a library to be used in the matured Scala toolchain. While the main emphasis of Gavial is to provide a usable web programming framework, a formal semantics specifying the exact behavior of its main APIs is available http://tzbob.be/gavial/semantics.pdf.

While Gavial has matured, we have plenty of future work in mind. The FFI already allows us to access external tools like databases, but we would prefer to have more APIs native to FRP. We would like to build an FRP database API. We imagine reading from the database using non-discrete behaviors (similar to reading from the DOM) and modifying database state using events.

**Acknowledgements**   Bob Reynders held an SB fellowship of the Research Foundation - Flanders (FWO) during this work. This work has been partly supported by the FWO-SBO Tearless project. This research was partly supported by the National Research Foundation of Korea grants from MoE (No. 2019R1I1A3A01058608).

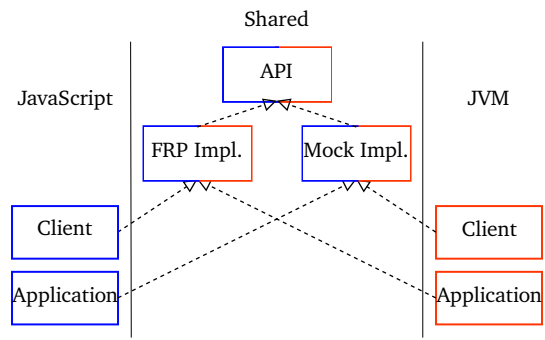

■ **Figure 7** Structure of the Gavial implementation.

## A Implementation: Gavial Architecture

There are several layers of implementation to Gavial as shown in figure 7. There are JavaScript (blue), JVM (red) and shared (blue & red) sections for their resp. platforms. A shared code section defines code that is included in both platforms.

At the top of Gavial there is an API definition that defines all primitives for a single tier, these include events, behaviors, tier-crossing primitives, etc. This definition has two non-platform specific implementations: FRP and Mock.

An FRP implementation of a tier implements the API with an FRP library. Events and behaviors actually work and cross-tier dependencies are passed through FRP primitives. A Mock implementation on the other hand implements nothing but multi-tier dependency tracking, the FRP primitives and their operations are essentially null-ops.

Both JVM and JavaScript libraries make use of the Mock and FRP tier. Note that the Client tier should do nothing on the JVM while the Application tier should not do anything in JavaScript. Both these tiers are implemented using the Mock tier while the FRP tier is used to implement the others.

## About the authors

**Bob Reynders** is a post doctoral researcher at Chonnam National University. Contact him at tzbobr@gmail.com.

**Frank Piessens** is a professor in the research group DistriNet (Distributed Systems and Computer Networks) at the Computer Science department of the Katholieke Universiteit Leuven. His main research interests are in the field of software security, where he focuses on the development of high-assurance techniques to deal with implementation-level software vulnerabilities and bugs, including techniques such as software verification, run-time monitoring, type systems, language based security and hardware-software co-design for security.

**Dominique Devriese** is an assistant professor in the Software Languages Lab at the VUB. His research interests are; formalising properties of object-oriented and object-capability programming languages, formal reasoning about capability machines (CPUs with a built-in form of low-level object capabilities), secure compilation and full abstraction properties and applying logical relations to prove them, and functional and dependently typed programming and programming languages, particularly Agda and Haskell